\documentclass[traditabstract]{aa}  

\usepackage{graphicx}
\usepackage{txfonts}
\usepackage{xcolor}
\usepackage{float}
\begin{document}

\title{Optical tomography of the born-again ejecta of A\,58}

   \author{B.~Montoro-Molina
          \inst{1}
          \and
          D.~Tafoya\inst{2} 
          \and
          M.~A.~Guerrero
          \inst{1}
          \and
          J.~A.~Toal\'{a}
          \inst{3}
          \and
          E. Santamar\'{i}a
          \inst{3}
          }

   \institute{Instituto de Astrof\'{i}sica de Andaluc\'{i}a, CSIC, Glorieta de la Astronom\'{i}a S/N, Granada, E-18008, Spain\\
         \email{borjamm@iaa.es}
         \and
         Department of Space, Earth, and Environment, Chalmers University of Technology, Onsala Space Observatory, SE-3992 Onsala, Sweden
        \and
        Instituto de Radioastronom\'{i}a y Astrof\'{i}sica, Universidad Nacional Aut\'{o}noma de M\'{e}xico, Antigua Carretera a P\'{a}tzcuaro 8701, Ex-Hda San Jos\'{e} de la Huerta, Morelia 58089, Michoac\'{a}n, Mexico
    }

   \date{\today}

 
\abstract
{In a born-again planetary nebula (PN), processed H-deficient material has been injected inside the old, H-rich nebula as a result of a very late thermal pulse (VLTP) event. 
Long-slit spectra have been used to unveil the chemical and physical differences between these two structures, but the ejection and shaping processes remain still unclear. 
In order to peer into the morpho-kinematics of the H-deficient ejecta in the born-again PN A\,58, we present the first integral field spectroscopic observations of a born-again PN as obtained with GTC MEGARA. 
We detect emission from the H$\alpha$, He~{\sc i}, [O\,{\sc iii}], [N\,{\sc ii}] and [S\,{\sc ii}] emission lines, which help us unveil the expansion patterns of the different structures. 
In combination with ALMA and {\it Hubble Space Telescope} data we are able to produce a complete view of the H-deficient ionized and molecular ejecta in A\,58. 
We propose an hourglass structure for the ionized material that embraces molecular high-velocity polar components, while bisected by an expanding toroidal molecular and dusty structure.
Our results leverage the role of a companion in shaping the VLTP ejecta in this born-again PN. 
}

   \maketitle
%

\section{Introduction}

Born-again planetary nebulae (PNe) are objects whose central stars (CSPNe) have experienced a very late thermal pulse (VLTP) when descending the white dwarf (WD) cooling track \citep{Schonberner1979,Iben1983}. 
These thermonuclear explosive events are rare, given that the VLTP is produced when the outer He layer of the WD reaches unique conditions to ignite into C, engulfing the remnant H on the surface. 
As a consequence, this process injects H-deficient material inside an old, H-rich PN, creating double-shell structure \citep[e.g.,][]{ToalaLora2021,Ary2023}.

Models predict that the duration of the VLTP is short \citep[10--200~yr;][]{Miller2006}, which makes it difficult to identify born-again PNe. This would explain that only a handful of objects have been identified as born-again PNe, being A\,30, A\,58, A78 and the Sakurai's Object the most studied cases \citep[see, e.g.,][]{Jacoby1979,Nakano1996,Clayton2006}. 
Spectroscopic studies have helped unveil the extreme abundances differences between the born-again ejecta and those of the outer H-rich nebula \citep{Jacoby1983,Hazard1980,Manchado1988,Montoro2022,Montoro2023,Wesson2008,Simpson2022}. 

The kinematics of the H-deficient ejecta is far from simple. 
High-dispersion optical spectroscopy works of the most evolved objects of this class, A\,30 and A\,78, found that the H-deficient material in these born-again PNe has velocities ranging from 40~km~s$^{-1}$ up to 500~km~s$^{-1}$ \citep{Meaburn1996,Meaburn1998}. 
In addition, {\it Hubble Space Telescope} ({\it HST}) observations show that the inner structures of the H-deficient ejecta in A\,30, A\,58 and A\,78 have bipolar morphology consisting of a disk-like (or toroidal) structure and a pair of bipolar ejections \citep{Borkowski1993,Borkowski1995,Clayton2013}. The H-deficient clumps have a tadpole (clump-head and tail) morphology, which is indicative of the complex interactions of this material with the photoionization flux and the current fast wind from the CSPN \citep{Fang2014,RodriguezGonzalez2022}.

The disk-jet morphology of the H-poor ejecta in born-again PNe is suggestive of the action of binary systems. 
Moreover born-again PNe are listed among the PNe with the highest abundance discrepancy factor, which is interpreted as caused by the evolution through a binary system \citep{Wesson2018}. 
The C/O abundances ratio of born-again ejecta seems also consistent with those of novae \citep{Lau2011}, although, once that the C trapped in dust is accounted for, the C/O ratio is actually more consistent with the predictions of a VLTP event of single stellar evolution models \citep[see][for the case of A\,30]{Toala2021}. 
Although VLTP events and binary systems would seem unrelated, a common envelope phase with a binary companion after the VLTP event has been recenlty invoked to explain the bipolar structures and their specific kinematic signatures \citep{RodriguezGonzalez2022}.

It is thus clear that an appropriate determination of the kinematics of the H-deficient ejecta is most needed to peer into the single versus binary scenarios. Thus far, the best determination for the youngest born-again PNe have been achieved studying the molecular emission. 
Atacama Large Milimeter/submillimeter Array (ALMA) observations of the Sakurai's Object and A\,58 have shown that in both cases the molecular CO emission exhibit bipolar outflows protruding from an expanding toroidal structure;  
in the Sakurai's Object the deprojected disk velocity is 53~km~s$^{-1}$ and that of the bipolar outflow  $\sim$1000~km~s$^{-1}$ \citep{Tafoya2023}, while in A\,58 the disk is estimated to expand at a velocity of 90~km~s$^{-1}$ and its bipolar outflow at 280 km~s$^{-1}$ \citep{Tafoya2022}.

\begin{table*}
\centering
\caption{Details of the GTC MEGARA observations of A\,58 analyzed in this work.}
\label{tbl:observation}
\setlength{\tabcolsep}{10pt} 
\renewcommand{\arraystretch}{1.1} 
\begin{tabular}{lcccccccc}
\hline
Dispersion Element      & R      & Spectral Range    & Exposure Time     & Moon   & Sky    & Airmass & Seeing \\
         &        & (\AA)             & (s)               &        &        &        &   (arcsec)    \\
\hline
VPH665HR-R  & 20050  & 6405.6--6798.0   & $3\times300$      & Dark   & Clear  & 1.49    & 0.9 \\
VPH481MR-B  & 13200  & 4585.7--5025.1   & $3\times600$      & Dark   & Clear  & 1.43    & 0.9   \\
VPH443MR-UB & 13050  & 4226.4--4625.8   & $6\times900$      & Dark   & Clear  & 1.57    & 0.9   \\
\hline
\end{tabular}
\end{table*}

In this paper we start a series of works to study the morpho-kinematics of the H-deficient ejecta of a sample of born-again PNe using the unrivaled capabilities of high-dispersion integral field spectroscopic (IFS) observations obtained with the Multi-Espectrógrafo en GTC de Alta resolución para Astronomía \citep[MEGARA;][]{GildePaz2018} at the 10.4 m Gran Telescopio Canarias (GTC). 
Here we present results for A\,58 that in conjunction with the available molecular emission detected by ALMA provide an unprecedented view of this born-again PN.

This paper is organized as follows. In Section~\ref{sec:obs} we describe our observations and their reduction. The analysis procedure of the data is presented in Section~\ref{sec:analysis}. The discussion of our results in presented in Section~\ref{sec:discussion}. Finally, our conclusions are presented in Section~\ref{sec:conclusions}.

\section{Observations and data Reduction}
\label{sec:obs}

\subsection{Integral Field Spectroscopy}

IFS observations of A\,58 were obtained on 2022 June 21 (Program ID 24-GTC29/22A) using MEGARA at the GTC of the Observatorio de El Roque de los Muchachos (ORM, La Palma, Spain). 
The Integral Field Unit (IFU) mode, also called Large Compact Bundle (LCB), was used.  
It consists of 567 hexagonal spaxels of 0\farcs62 in diameter resulting in a field of view (FoV) of 12\farcs5$\times$11\farcs3. 
The volume phase holographics VPH443-MR (MR-UB), VPH481-MR (MR-B), and VPH665-HR (HR-R) were used as dispersion elements.  
The details of the observations, including spectral properties, exposure times, and observing conditions, are presented in Table~\ref{tbl:observation}.

The raw MEGARA data were reduced following the Data Reduction Cookbook \citep{Pascual2019} using the \textit{megaradrp} v0.10.1 pipeline released on 2019 June 29. 
This pipeline applies sky and bias subtraction, flat-field correction using halogen internal lamps, 
wavelength calibration, and spectra tracing and extraction.
The final output is a FITS file that contains the science-calibrated row-stack-spectra (RSS) for each fiber, with metadata of the relative fiber positions to the IFU center. 
This RSS FITS file is converted into a $52\times 58$ map of 0.2 arcsec pix$^{-1}$ on the spatial dimension and 4300 spaxel along the spectral axis using the regularization grid task \emph{megararss2cube}. 
The flux calibrations were performed using observations obtained immediately after those of A\,58 of the spectrophotometric standard stars HR\,7950, HR 7596, and HR 4963 for the VPH665-HR, VPH481-MR, and VPH443-MR, respectively.

\subsection{Long-slit Echelle Spectroscopy}

High-dispersion spectroscopic observations of A\,58 were obtained on 2002 June 23 using the echelle spectrograph on the Cerro Tololo Interamerican Observatory (CTIO) 4 m V\'\i ctor Blanco (a.k.a. Blanco) telescope. 
The spectrograph was used in its long-slit mode with the 6563/75 narrow-band filter, whose $\approx$75 \AA\ in FWHM bandwidth isolates the echelle order including the H$\alpha$ and [N~{\sc ii}] $\lambda\lambda$6548,6584 emission lines. 
The 79 line~mm$^{-1}$ echelle grating and the long-focus red camera were used, resulting in a reciprocal dispersion of 3.4 \AA~mm$^{-1}$. 
The data were recorded with the SITe 2K CCD \#6, whose pixel size of 24 $\mu$m provides a spatial scale of 0\farcs26 and a spectral sampling of 0.081 \AA~pixel$^{-1}$ (i.e., 3.7 km~s$^{-1}$~pixel$^{-1}$) along the dispersion direction. 
The slit has an unvignetted length of 3 arcmin and its width was set to 1\farcs4, resulting in an instrumental resolution of 9.1 km~s$^{-1}$. 
Two individual 750 s exposures were obtained with the slit oriented along a position angle (PA) of 50$^\circ$, i.e., along the central ejecta.  
The angular resolution, determined by the seeing measured at the DIMM, was $\approx$1\farcs0.

The spectra were reduced using standard IRAF \citep{Tody1993} routines for two-dimensional spectra. 
The wavelength scale and geometrical distortion were corrected using a two-dimensional fit to an arc exposure obtained using Th-Ar calibration lamps immediately after the science exposure. 
The deviation of the residuals of the two-dimensional fit to the Th-Ar arc is found to be better than $\approx$0.004~\AA\ (0.2~km~s$^{-1}$). 
The telluric emission lines, which includes mostly OH emission lines, but also the geocoronal H$\alpha$ line, were removed by fitting and subtracting the background using low-order polynoms. 
Before this procedure, the telluric lines were used to confirm the accuracy of the wavelength calibration to be better than 0.3~km~s$^{-1}$ using their theoretical wavelengths \citep{Osterbrock1996}.

\subsection{Milimeter/submillimeter Interferometric Observations}

We retrieve high angular resolution (0\farcs07 $\times$ 0\farcs1) observations obtained with the ALMA used to detect the continuum and molecular emission from A\,58. These observations correspond to project 2019.1.01408.S (PI: D. Tafoya) and their details can be found in \cite{Tafoya2022}.

\begin{figure*}
\centering
\includegraphics[width=1\linewidth]{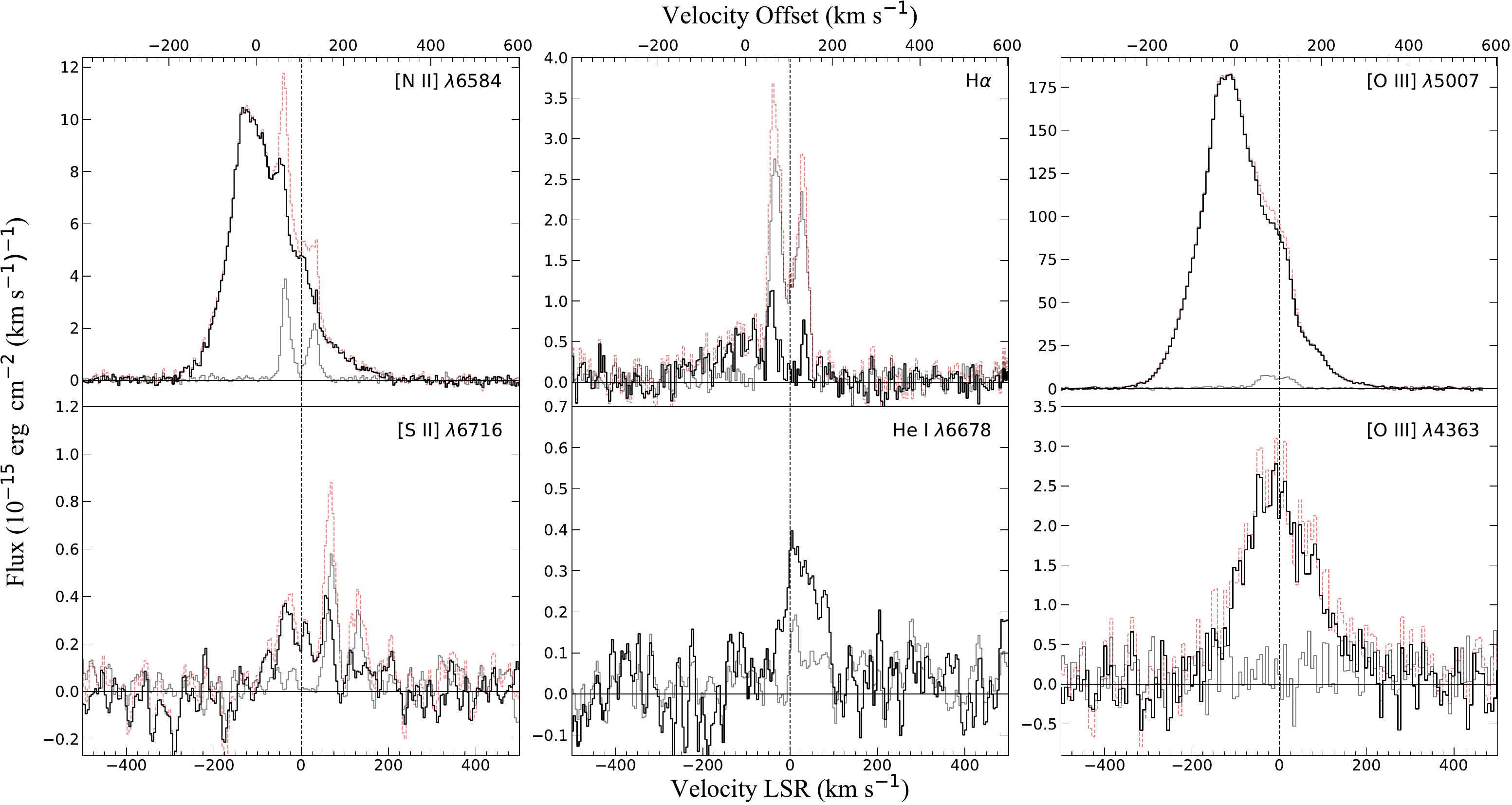}
\caption{
GTC MEGARA emission line profiles of the central ejecta of A\,58 (red dashed line) extracted from a 1\farcs8 in radius circular region, and surrounding nebula (grey solid line) averaged from four regions of the same size located at the corners of the MEGARA FoV.  
The [N~{\sc ii}] $\lambda$6584, H$\alpha$, [O~{\sc iii}] $\lambda$5007, and [S~{\sc ii}] $\lambda$6716 emission line profiles of the central ejecta include notable contribution from the surrounding nebula.  
The surrounding nebula emission profiles of these lines have been subsequently subtracted to obtain the net emission line profiles of the central ejecta (black solid line), whereas for the He~{\sc i} $\lambda$6678 and [O~{\sc iii}] $\lambda$4363 emission line profiles only a constant level representative of the continuum emission has been subtracted.  
As described in the text the subtraction of the emission from the surrounding nebula is not perfect most likely given its non uniform surface brightness.
The horizontal line marks the continuum level, while the vertical line marks the systemic velocity of +103 km~s$^{-1}$ of the surrounding nebula. 
}
\label{fig:lines_profiles}
\end{figure*}

\section{Data analysis}
\label{sec:analysis}

\subsection{Spectral Line Profiles}

The GTC MEGARA observations detected emission from the [O~{\sc iii}]$\lambda\lambda4363,4959,5007$, 
[N~{\sc ii}]$\lambda\lambda6548,6584$, 
[S~{\sc ii}]$\lambda\lambda6716,6731$, {He~\sc i} $\lambda6678$, {He~\sc ii} $\lambda4686$, H$\beta$  and H$\alpha$ emission lines. 
We note that the C~{\sc ii} 4267~\AA\ emission line is not detected in the VPH433-MR observations at a 3-$\sigma$ upper limit of 3.6$\times$10$^{-16}$ erg~cm$^{-2}$~s$^{-1}$.

Spectral profiles of key emission lines of A\,58 are presented in Fig.~\ref{fig:lines_profiles}. 
The profiles of the H-poor ejecta (red dashed histograms in Fig.~\ref{fig:lines_profiles}) have been extracted from a circular region 1\farcs8 in radius around the brightest region. 
These spectra include significant contamination from the surrounding old, H-rich nebula.  
This emission is estimated by averaging four apertures located in each quadrant of MEGARA's FoV with the same size as that used for the central ejecta.  
The emission line profiles from the surrounding nebula (gray solid histograms in Fig.~\ref{fig:lines_profiles}) present a well-marked double-peak structure in the [N~{\sc ii}] and H~{\sc i} Balmer lines, as well as in the fainter and noisier [S~{\sc ii}] emission lines.  
The [O~{\sc iii}] emission line seems to be consistent with a double-peak profile, but the components are broader and blended.  
Although this could be attributed to the lower spectral resolution of the blue MR-B VPH ($R \simeq 13200$) that registered this line compared to the red HR-R VPH ($R \simeq 20050$), the similarity of the H$\beta$ (not shown here) and H$\alpha$ nebular profiles implies that the components of the nebular [O~{\sc iii}] emission line are truly broader. 
Finally the [O~{\sc iii}] $\lambda4363$ and He lines do not present emission from the surrounding old nebula. 

The blue component of the different emission lines from the outer nebula is consistently narrower and more intense than its red counterpart. 
If the expansion velocity of the nebula, $V_{\mathrm{exp}}$, is assumed to be half the separation between the red and blue components, we derive velocities of 30 and 35 km~s$^{-1}$ for H$\alpha$ and [N~{\sc ii}], respectively, 
which are consistent with the expansion velocities of these emission lines derived from the CTIO long-slit echelle data.  

\begin{figure*}
\centering
\includegraphics[width=0.9\linewidth]{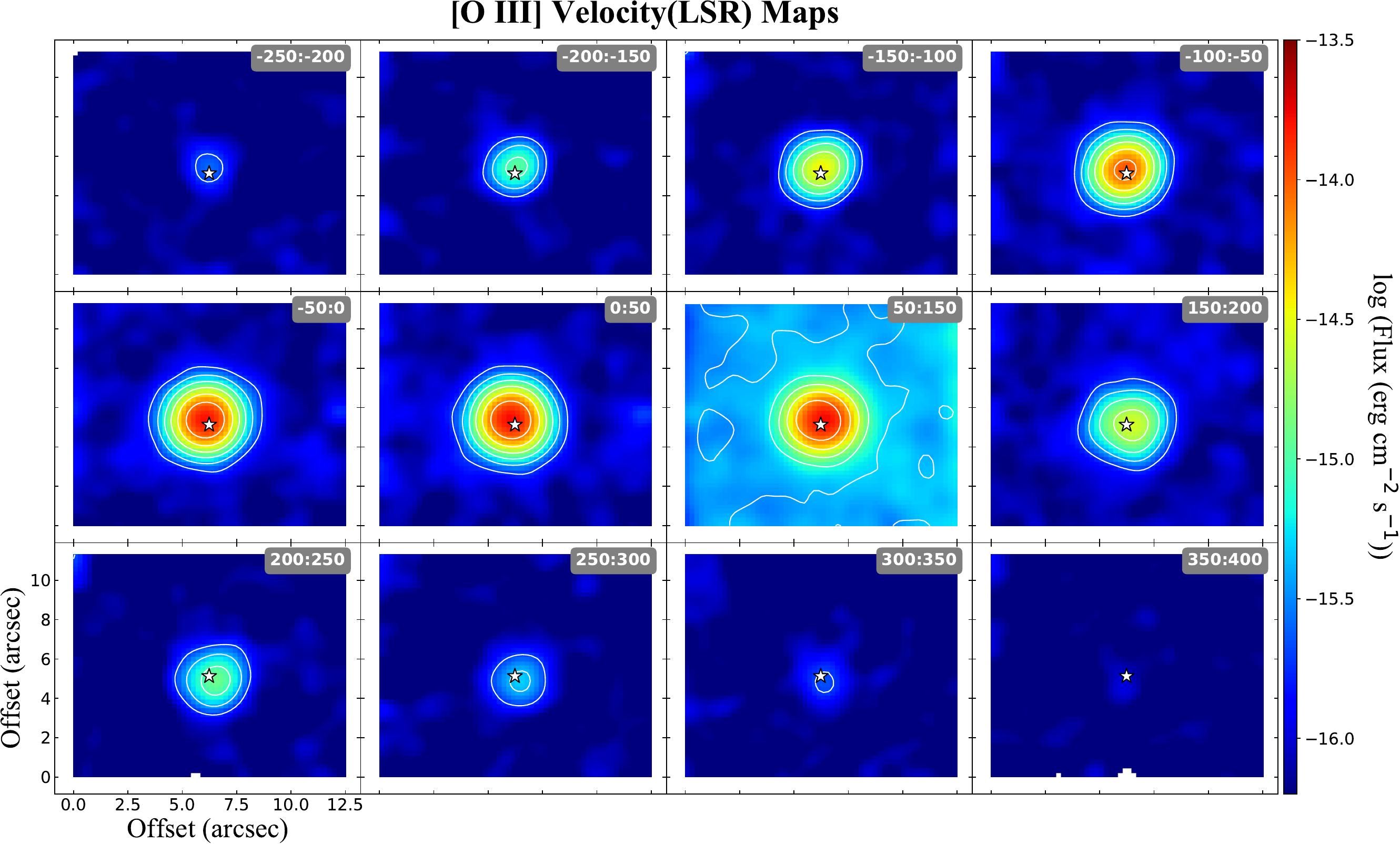}\\
\includegraphics[width=0.9\linewidth]{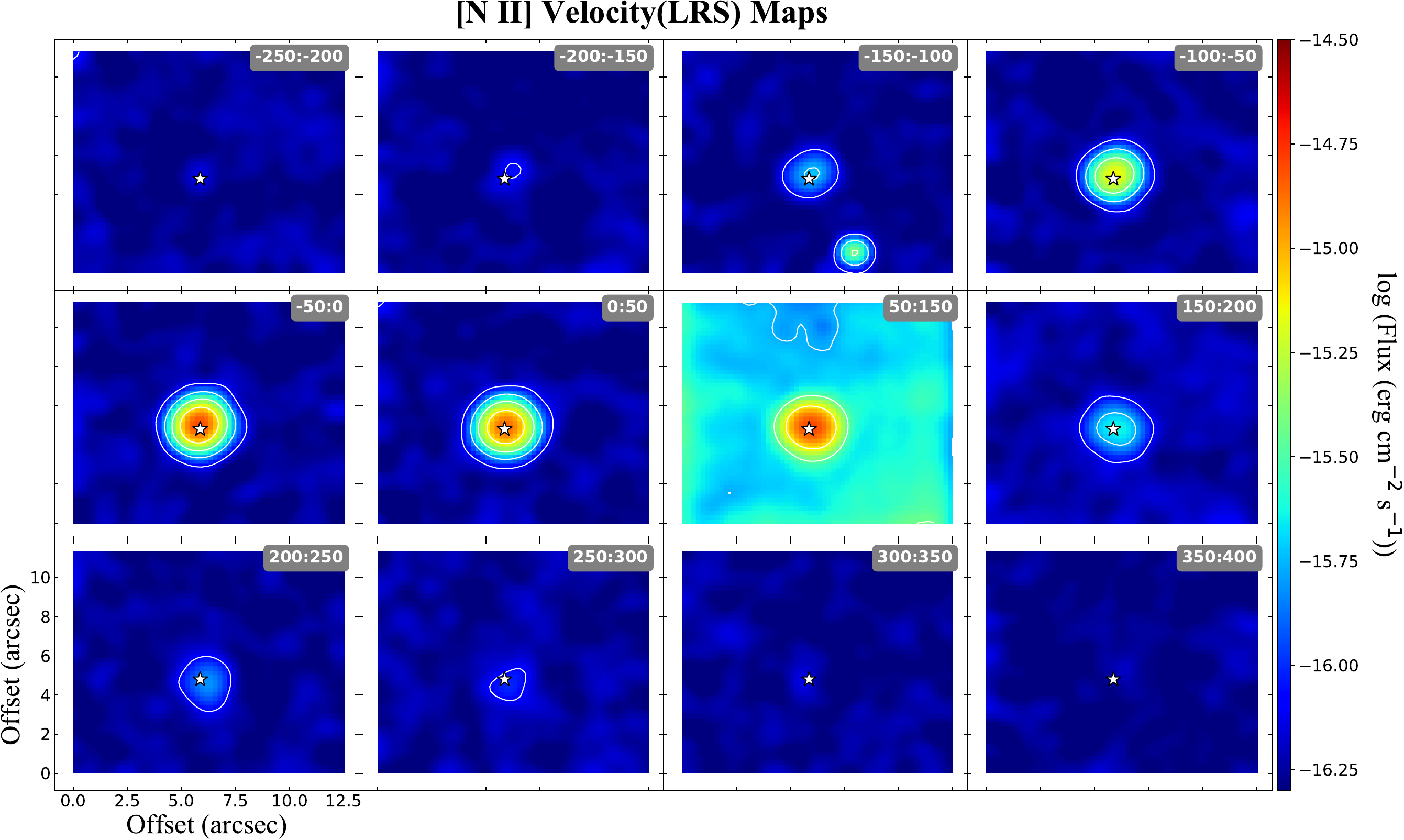}
\caption{GTC MEGARA VPH481-MR [O~{\sc iii}] $\lambda$5007 (top) and VPH665-HR [N~{\sc ii}] $\lambda$6584 (bottom) channel maps of A\,58.
Each map shows the integrated emission in the channles within the LSR velocity range labeled on the upper right corner of each map.
Contours corresponding to
2.4\%, 5\%, 11\%, 23\%, and 48\% of the emission peak (1.71$\times10^{-14}$ and $1.61 \times10^{-15}$ for [O~{\sc iii}] and [N~{\sc ii}], respectively) are overlaid in all maps (solid white lines). 
The approximate location of the central star is marked by a white star.}
\label{fig:vel_maps}
\end{figure*}

Once the contribution of the old surrounding nebula is subtracted from the spectrum of the aperture encompassing the central ejecta, the remaining emission shall be attributed to net emission from that central ejecta (solid black lines in Fig~\ref{fig:lines_profiles}).  
We note, however, that some residual emission subsists, as, for instance, the blue-shifted peak at $V_{\rm LSR} \approx 60$ km~s$^{-1}$ in the [N~{\sc ii}] emission line and the narrow components at $V_{\rm LSR} \approx 60$ km~s$^{-1}$ and $\approx 130$ km~s$^{-1}$ in H$\alpha$.  
These residuals most likely result from spatial variations of the emission of the surrounding old nebula \citep[see, e.g., figure 2 in][]{Guerrero1996}. 

The spectral profile from the central ejecta is asymmetric in all emission lines shown in Fig.~\ref{fig:lines_profiles}. 
The emission peaks bluewards, very notably in the [N~{\sc ii}] and [O~{\sc iii}] emission lines at $V_{\rm LSR} \approx -20$ km~s$^{-1}$.  
In these emission lines, the line profiles show wings that extends in the range $-200$ km~s$^{-1} \lesssim V_{\rm LSR} \lesssim +300$ km~s$^{-1}$. 
The overall properties of these emission lines are very similar to those presented by \cite{Pollacco1992}, although the data presented here have higher signal-to-noise ratio and reveal more clearly the line wings.  
Otherwise only the blue component is detected in the H$\alpha$, [S~{\sc ii}], and He~{\sc i} emission lines.  
We emphasize that the non-detection of the He~{\sc ii} $\lambda$4686 emission line discards the possible contribution of He~{\sc ii} $\lambda$6560 to the emission bluewards of H$\alpha$, which is thus confidently attributed to the ejecta.  
This is the first time that the H$\alpha$ emission of the born-again ejecta is unambiguously detected, although a revision of figure~2 in \cite{Pollacco1992} reveals arguable evidence for its detection.

\subsubsection{Systemic velocities of A\,58 and its central ejecta}

Considering the average of the nebular red and blue components to be the systemic velocity, $V_\mathrm{sys}$, we derive velocities in the Local Standard of Rest (LSR) from the CTIO echelle data of 101.8 km~s$^{-1}$ and 104.4 km~s$^{-1}$ for the H$\alpha$ and [N~{\sc ii}] lines, respectively. 
An average of 103 km~s$^{-1}$ can thus be adopted for the systemic velocity of the old surrounding nebula, which is marked by a vertical dashed line in the different panels of Fig.~\ref{fig:lines_profiles}. 
These values are found to be consistent to the observed dips in the GTC MEGARA H$\alpha$ and [N~{\sc ii}] $\lambda$6584 spectral profiles of the outer nebular shell shown in Fig.~\ref{fig:lines_profiles}.

The only previous high-dispersion spectra of A\,58, obtained with the Anglo-Australian Telescope Coud\'e \'echelle spectrograph (UCLES), implied a systemic velocity in the LSR of its outer nebula of 86 km s$^{-1}$ \citep[corresponding to a heliocentric velocity of 70~km~s$^{-1}$,][]{Pollacco1992}. 
This value is not associated to a specific emission line by the authors, neither there is a description of the slit position, besides the information on the 2$^{\prime\prime}$-wide slit.  
Due to the broad slit width, the nebular emission in the H$\alpha$ and [N~{\sc ii}] emission lines is more prominent than the profile shown in Fig.~\ref{fig:lines_profiles}, making the double-peak quite noticeable. 
Later on, \citet{Clayton2013} analyzed the spectra presented by \citet{Pollacco1992} and derived a systemic velocity in the LSR of 96 km~s$^{-1}$ (or a heliocentric velocity of 80 km~s$^{-1}$).

The differences between the systemic velocity derived here and those reported by \citet{Pollacco1992} and \citet{Clayton2013} are 17~km~s$^{-1}$ and 7~km~s$^{-1}$, respectively. 
It is important to note that the original spectra presented by \citet{Pollacco1992} were not available to \citet{Clayton2013}, who used a digitized version of the figures of the spectra and therefore adopted the same wavelength calibration. 
If we compare the dispersion of the UCLES and CTIO spectra, the spectral dispersion of the latter is approximately ten times higher than that of the UCLES spectra. 
The calibration and analysis of the CTIO data seems very reliable, as described in Section~2.2.  
Therefore the value of 103 km~s$^{-1}$ presented here is preferred for the systemic velocity of the outer shell of A\,58. 

Regarding the central ejecta of A\,58, it is not possible to derive a systemic velocity from the optical emission lines as 
their profiles are highly asymmetric, most likely due to the strong extinction of this region \citep{Montoro2022} absorbing preferentially the red component. 
On the other hand the radio emission of molecular material, which is mostly unaffected by the extinction, allows a more reliable determination of the systemic velocity of the recent ejecta. 
The spectral profile of the CO ($J=3\rightarrow2$) emission from the central region of A\,58 detected by the 12 m Atacama Pathfinder Experiment (APEX) implied an average velocity in the LSR of 96$\pm$11 km~s$^{-1}$, which was interpreted as the systemic velocity of the molecular gas around V605\,Aql \citep{Tafoya2017} in agreement with the value proposed by \citet{Clayton2013}.  
We note, however, that the quality of the APEX spectral profile of the CO ($J=3\rightarrow$2) emission line is limited.  
The subsequent ALMA spectral profiles of the CO ($J=3\rightarrow$2), HCN ($J=4\rightarrow$3), and HCO$^+$ ($J=4\rightarrow$3) emission line profiles presented by \citet{Tafoya2022} are of higher quality, but these were not used to derive new values of the systemic velocity.  
Instead, the authors conclude that the systemic velocity derived from APEX were consistent with the double-peak symmetric spectral profiles detected in the ALMA observations.

There is thus a difference of 7 km~s$^{-1}$ between the systemic velocity of the H-rich outer nebula derived from the CTIO optical data and that of the H-poor born-again ejecta derived from the APEX radio data, but these can be considered consistent given the uncertainties. 

Indeed the dips between the blue and red components of the CO ($J=3\rightarrow$2) and HCO$^+$ ($J=4\rightarrow$3) emission lines detected by ALMA are suggestive of a value of the systemic velocity slightly larger than that derived from APEX data. 
We will hereafter consider the systemic velocity of A\,58 and the H-poor ejecta around V605\,Aql to be 103 km~s$^{-1}$. 
For this systemic velocity, the peaks of the H-poor ejecta [N~{\sc ii}] and [O~{\sc iii}] emission line profiles are shifted by $\approx -120$ km~s$^{-1}$ with respect to the systemic velocity, and their wings extend $\approx -300$ km~s$^{-1}$ and $\approx +200$ km~s$^{-1}$.

\subsection{Channel Maps}

The GTC MEGARA data cube also allows obtaining spatial information for each spectral channel. 
The channel maps of the bright [O~{\sc iii}] $\lambda5007$ and [N~{\sc ii}] $\lambda6584$ emission lines in the central ejecta of A\,58 are shown in Fig.~ \ref{fig:vel_maps}. 
Each map is obtained by combining channels within a velocity range of 50~km~s$^{-1}$, except those around the systemic velocity including the nebular emission that have been computed for a velocity range of 100~km~s$^{-1}$.

Fig.~\ref{fig:vel_maps} reveals a trend in the spatial location of the emission from the born-again ejecta of A\,58.   
The bluest velocity channels, in the range from $-250$ to $-50$ km~s$^{-1}$, show emission Northeast of the CSPN,  
whereas the emission of the reddest velocity channels in the range from $+50$ to $+350$ km~s$^{-1}$ departs notably towards the Southwest.

\subsubsection{Relative location of V605\,Aql and the optical outflow}

The CSPN of A58, V605 Aql, has not been visible since 1923, when it underwent a brightening to fade away immediately afterwards \citep{Seitter1985, Harrison1996}. 
Ninety years later, \cite{Clayton2013} estimated its position at  R.A.=19$^{\mathrm{h}}$18$^{\mathrm{m}}$20$^{\mathrm{s}}$.538, DEC.=+1°46$^\prime$58\farcs74 under the reasonable assumption that the {\it HST} F547M image was dominated by the C~{\sc iv} Wolf–Rayet emission feature in the spectrum of V605\,Aql, therefore its peak emission revealing its position. 
The accuracy of these coordinates, however, is affected by the intrinsic limitations of \emph{HST} to assign absolute positions.  
Indeed the comparison of the \emph{HST} coordinates of V605\,Aql with the position of the CO molecular emission derived from ALMA observations, which have higher absolute precision in establishing coordinates, revealed a notorious displacement between them.

At any rate there is no information on the absolute positioning of the MEGARA data cubes, which compromises a comparison of the spatial properties of the optical outflow with the \emph{HST} and ALMA observations. 
To compensate this lack of information and to allow the spatial comparison of the \emph{HST}, ALMA and MEGARA observations, the stellar continuum in the MEGARA observations can be used to determine the position of V605\,Aql.  
Continuum images were thus built from the VPH481-MR and VPH665-HR data cubes collapsing spectrally all channels with no line emission contribution, avoiding the noisiest ones as well.  
The position of V605\,Aql was then determined by adjusting a two-dimensional Gaussian to the stellar emission located near the center of the image. 
This procedure revealed an offset between the position of the star in the VPH481-MR and VPH665-HR data cubes of 1.3 pixels, i.e., 0\farcs26, which is most likely due to irregularities in the surface and/or exact positioning of these two VPHs.  

It is worth noting that we also explored the possibility of assigning the position of the centroid of the emission at the systemic velocity to V605\,Aql.   
However, this method revealed a quite noticeable shift with the position of the stellar continuum.  
Apparently the emission at the systemic velocity does not correspond to the location of the CSPN, which reveals a far from simple 3D physical structure of the ejecta.

\subsection{Boosting the Tomography Spatial Resolution}
\label{subsec:centroids}

\begin{figure*}
\centering
\includegraphics[width=0.46\linewidth]{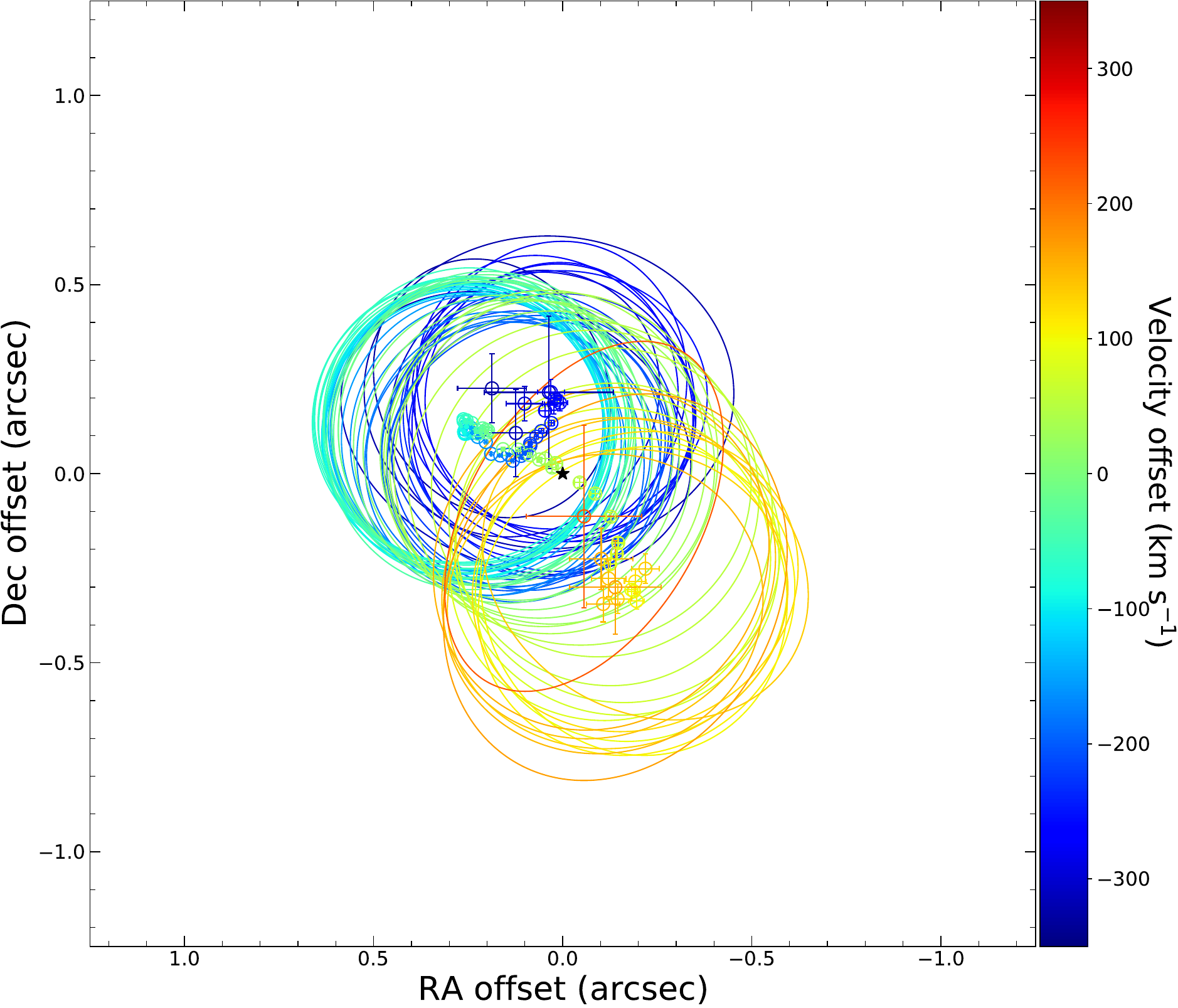}
\includegraphics[width=0.46\linewidth]{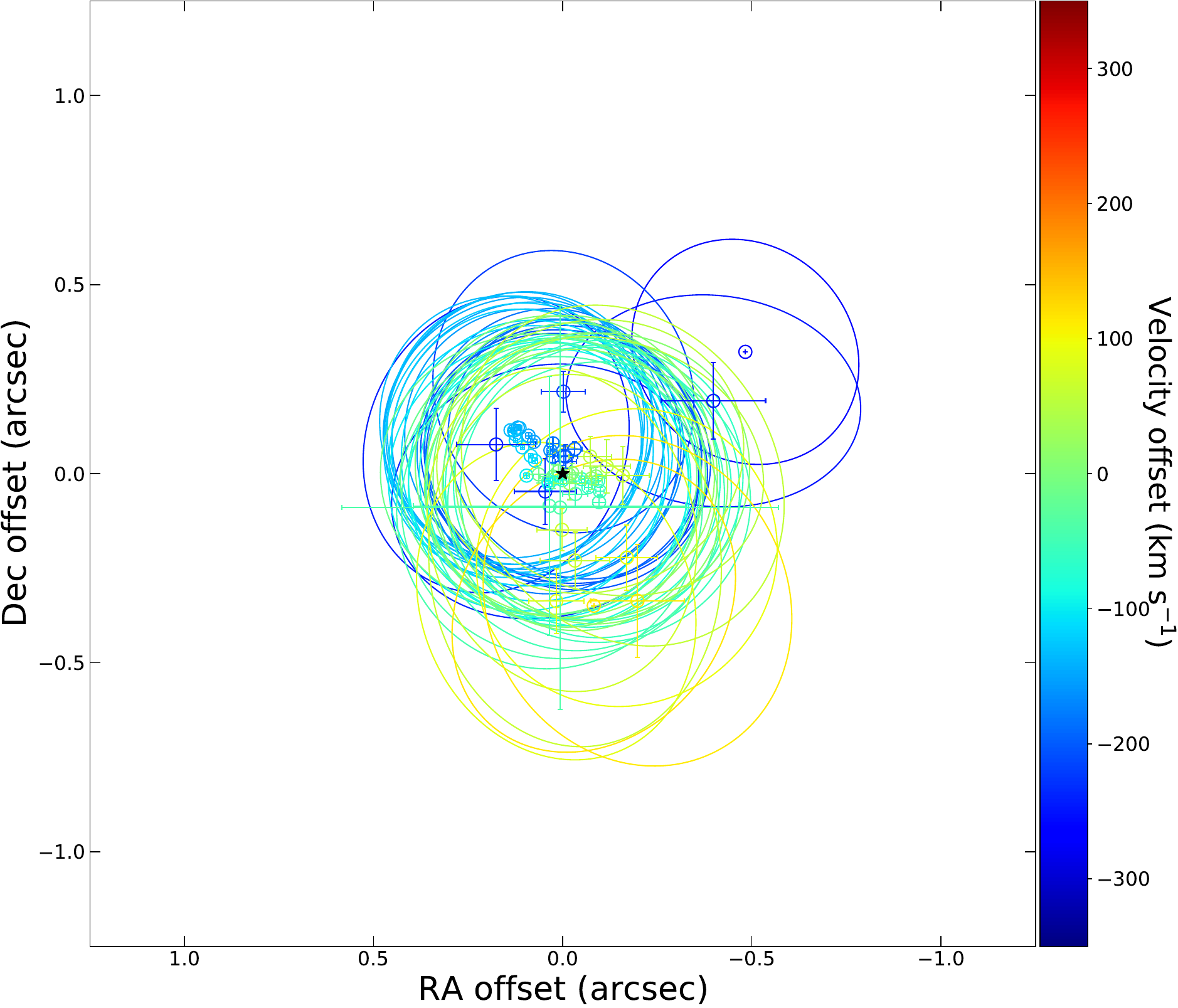}
\\
\includegraphics[width=0.46\linewidth]{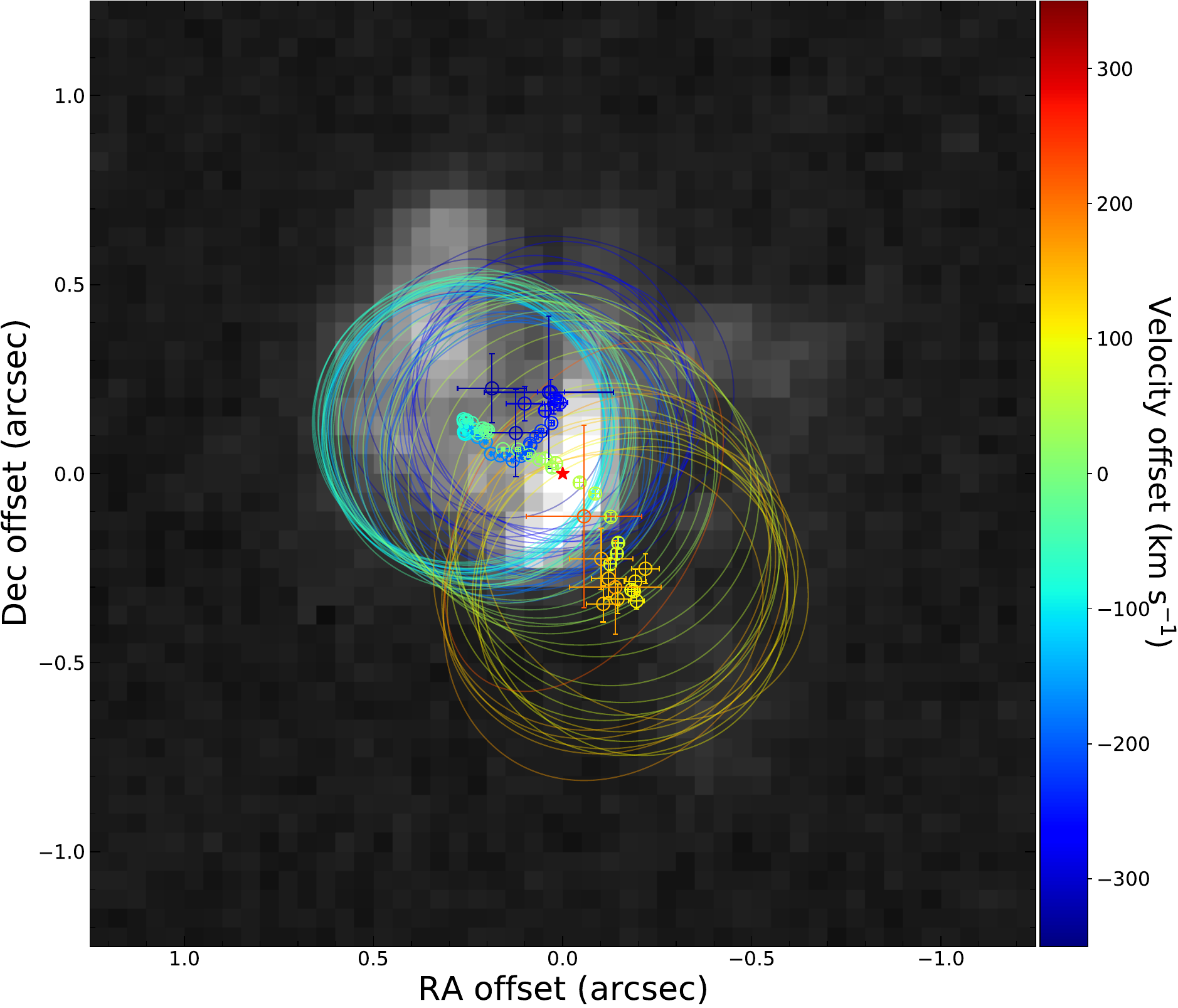}
\includegraphics[width=0.46\linewidth]{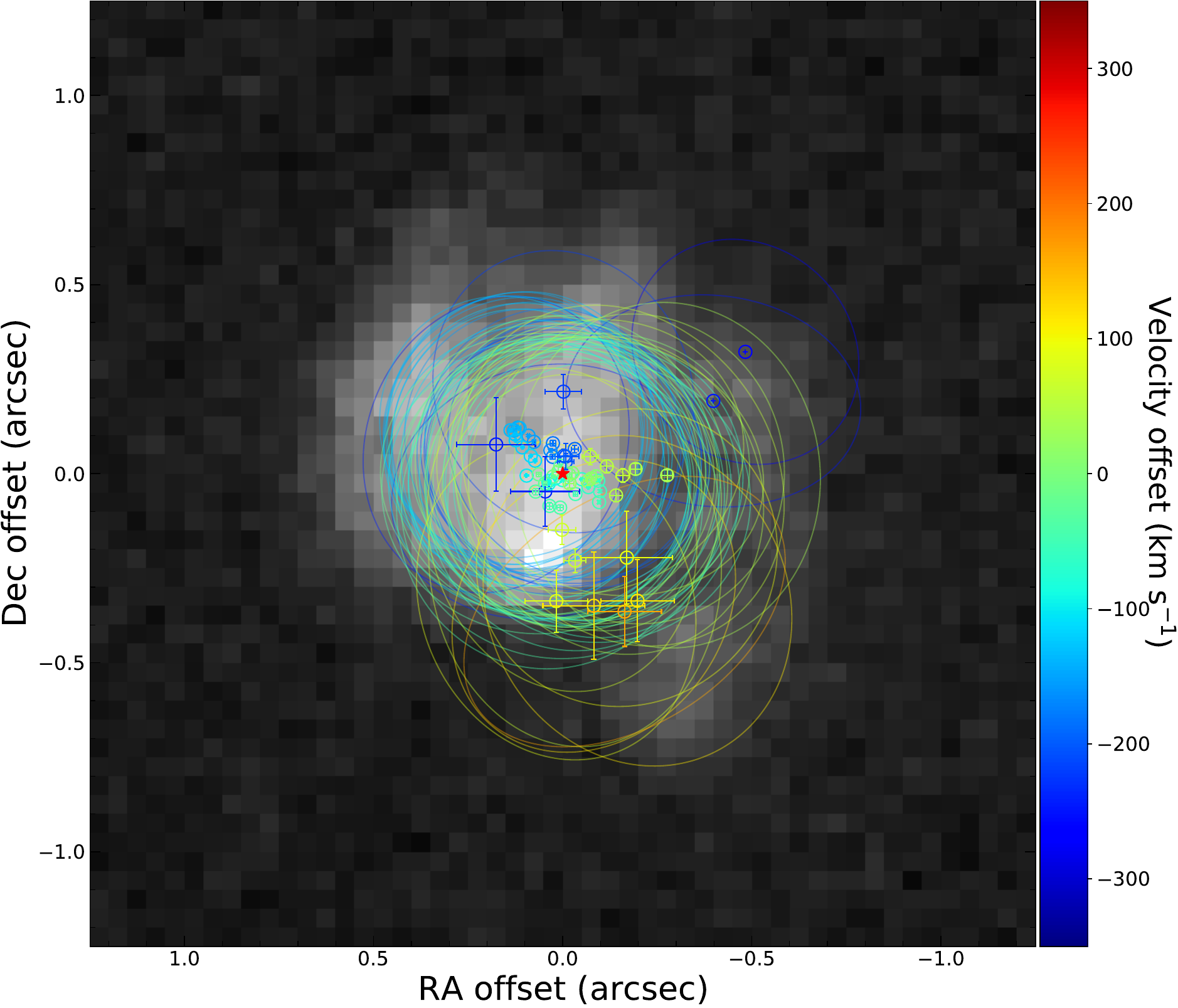}
\\
\includegraphics[width=0.46\linewidth]{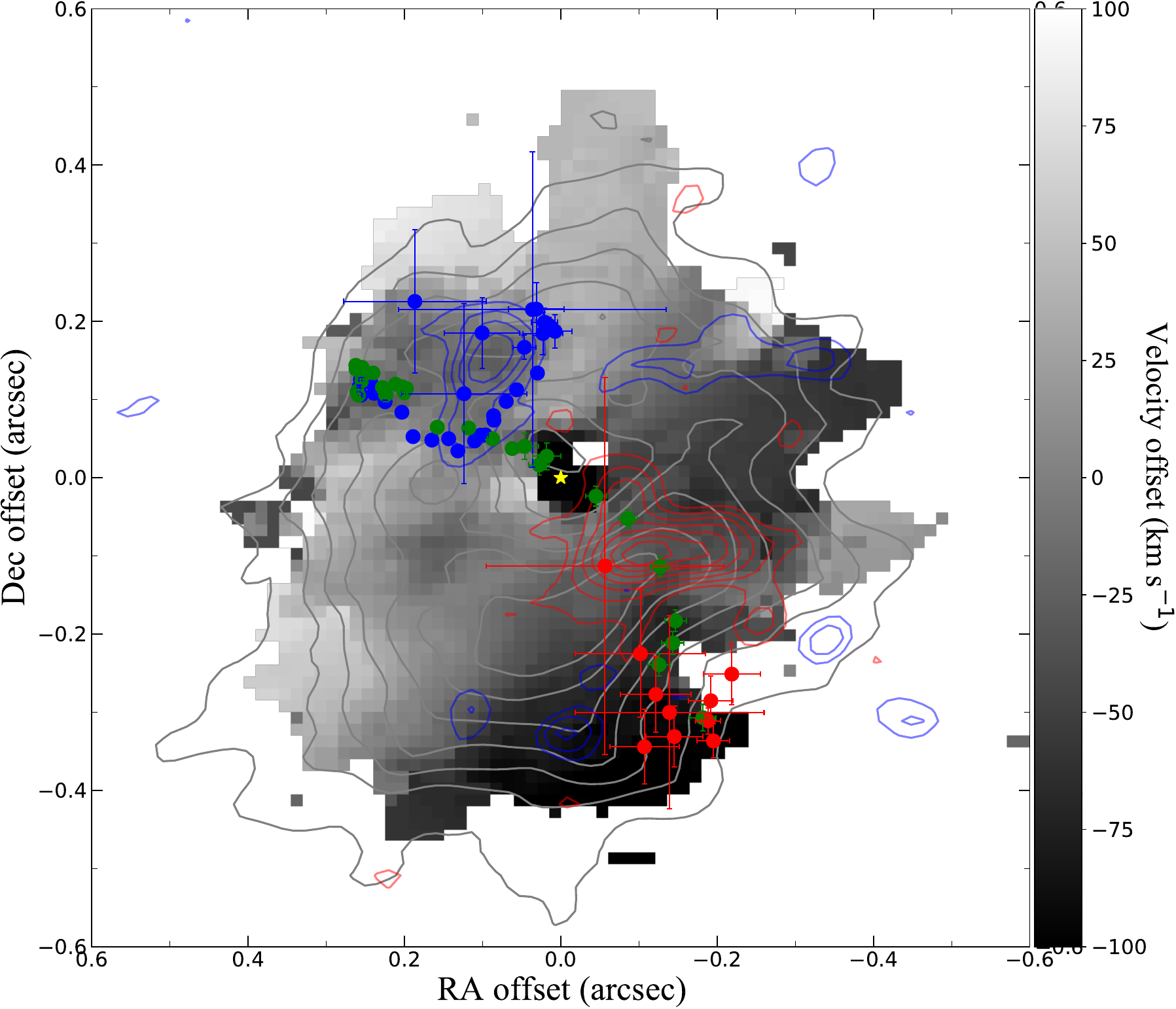}
\includegraphics[width=0.46\linewidth]{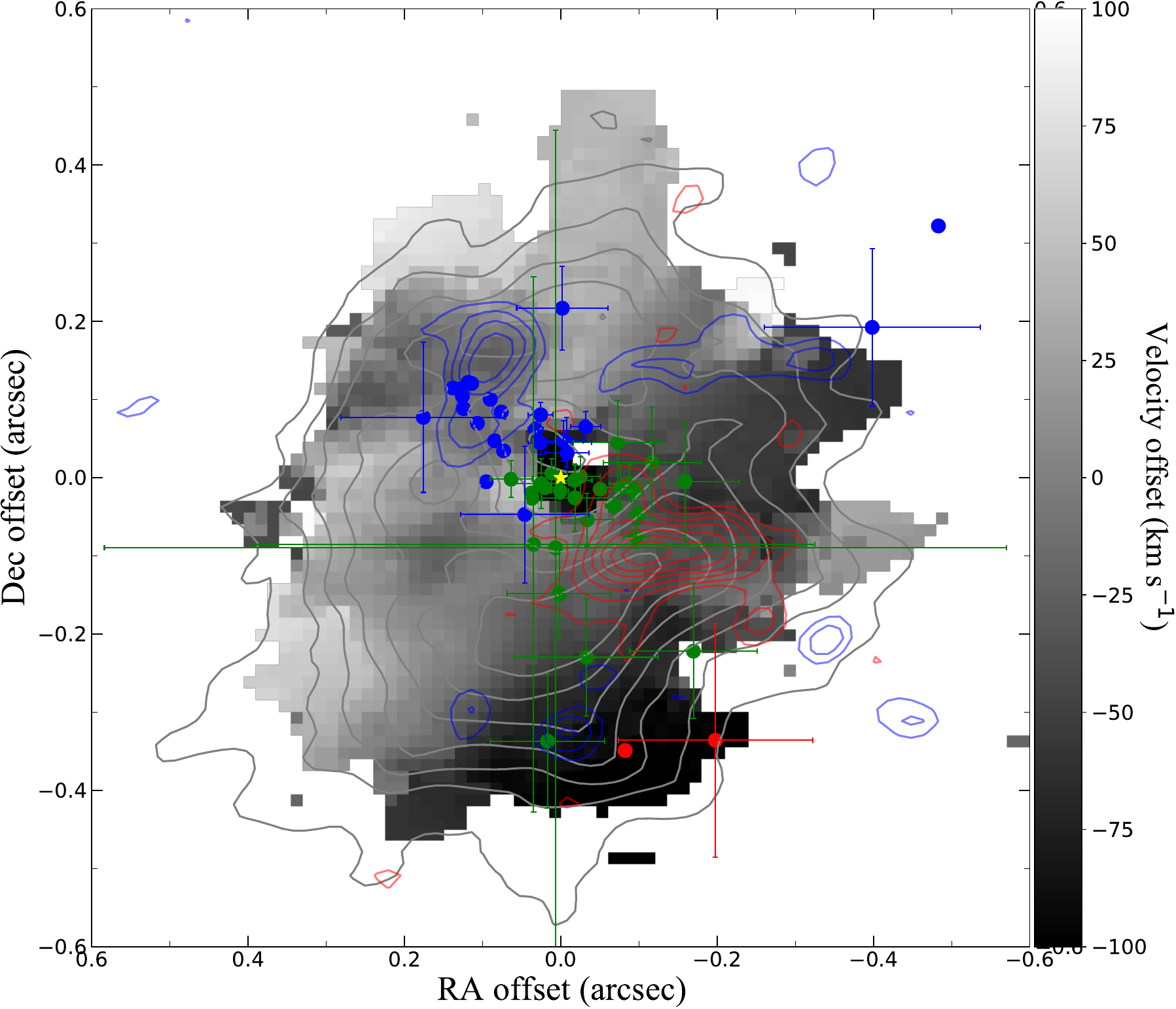}
\caption{
Top panels shows the positions of the centroids and the FWHM (ellipses contours) of 2-dimensional Gaussian fits to the emission of [O~{\sc iii}] $\lambda5007$ (left panels) and [N~{\sc ii}] $\lambda6584$ (right panels) at each spectral channel of the outflow of A\,58 with a color-coded according to their velocity with respect to the $V_\mathrm{sys}$ (see text for details).  
The middle panel show the top panels superimposed over the \emph{HST} [O~{\sc iii}] $\lambda5007$ (left) and [N~{\sc ii}] $\lambda6584$ (right) images of A\,58 obtained in 2009 (PI: G. Clayton; Program ID: 11985). 
The bottom panel present the spatio-kinematic distribution of CO ($J=3\rightarrow2)$ obtained with ALMA \citep[see figure 5 in][]{Tafoya2022}. 
In these panels the centroids in the velocity range  $-$100 km~s$^{-1}$ $\simeq$ V$_{\mathrm{offset}}$ $\simeq$+100 km s$^{-1}$ are plotted in green whereas centroids with velocities $< -100$ km~s$^{-1}$ and $> +100$ km~s$^{-1}$ are represented by blue and red filled dots, respectively. 
The approximate location of the central star is marked by a black, red and yellow stars in the top, middle and bottom panels, respectively.
}
\label{fig:centroids} 
\end{figure*}

The spatial resolution of the GTC MEGARA data is limited by the $\approx$0\farcs9 seeing during the observations and the coarse 0\farcs62 in diameter spaxel sampling, which results in a spatial resolution $\approx$1\farcs1.  
Otherwise the spectral resolution allows investigating the spatial location of the born-again ejecta at different velocities.  
The high S/N achieved in the MEGARA data actually allows increasing the accuracy of the spatial location of the emission at each velocity channel by a factor $\approx$10 by the determining the position of the centroid of the optical emission through a two-dimensional Gaussian fit \citep{Condon1997}.

In the top panel of Fig.~\ref{fig:centroids} we present both the centroids and the FWHMs (represented as ellipses) of the Gaussian resulting from the fittings for the [O~{{\sc iii}] (left) and [N~{{\sc ii}] (right) emission lines, with the position of V605\,Aql marked using a $\star$ symbol. 
These fittings were conducted for channels within the LSR velocity range from $-250$ to $+350$~km~s$^{-1}$. 
Channels with associated centroids located more than 1$^{\prime\prime}$ away from the position of V605\,Aql or with Gaussian FWHM exceeding 1\farcs5 in any of its axes were excluded. 
The centroids and FWHMs are represented with a color code indicating their velocity with respect to the adopted systemic velocity of +103~km~s$^{-1}$.

The spatial distribution of the centroids provides a clearer representation of the trend hinted in Fig.~\ref{fig:vel_maps}. 
The red-shifted emission shifts notably towards the southeast (SE) direction, while the blue-shifted emission moves towards the northeast (NE) direction, which is more pronounced for [O~{\sc iii}] than for [N~{\sc ii}]. 
Furthermore, the [O~{\sc iii}] emission exhibits a much more neat structure than that of [N~{\sc ii}], with centroids of velocity channels ranging from 
$-200$ to $+80$ km~s$^{-1}$ with respect to the systemic velocity consistently oriented along a specific direction, whereas the centroids of the [N~{\sc ii}] emission are more widely scattered, covering a slightly broader region around the central star. 
This dispersion cannot be completely attributed to the lower S/N ratio of this line, 
but more likely to a broader spatial distribution of the [N~{\sc ii}] emission, may be located outside that of [O~{\sc iii}].

The spatial behavior of the [O~{\sc iii}] emission of the highest velocity channels is noteworthy. 
The centroids trace a peculiar loop, moving back towards the position of the CSPN first, shifting then to the northwest (NW) direction, and finally receding back to a lesser extent to the northeast (NE) direction.  
Similar trend is not observed for the red-shift emission, which seems to simply move away with increasing velocity. 
This is further discussed in Section~\ref{sec:discussion}.

The middle panels of Fig.~\ref{fig:centroids}} show the {\it HST} [O~{\sc iii}] and [N~{\sc ii}] images with the MEGARA centroids and FWHMs of the corresponding top panel superimposed for comparison. 
The location of the CSPN was set at the peak of the [O~{\sc iii}] image as settled by \citet{Clayton2013} based on the \emph{HST} F547M continuum image.
The emission observed from the ejecta by MEGARA is consistent with the emission from the {\it HST} images, although the emission in the MEGARA maps appears to cover a slightly smaller region than in \emph{HST}.  
This is somehow unexpected because the expansion of the ejecta $\simeq$10 mas~yr$^{-1}$ \citep{Clayton2013} implies an even larger size of the emission from the time lapse $\approx$13.3 years between the \emph{HST} images (March 2009) and the MEGARA observations (June 2022).

Finally the lower panels of Fig.~\ref{fig:centroids} show the CO ($J=3\rightarrow$2) first momentum image presented by \cite{Tafoya2022} overplotted by the MEGARA [O~{\sc iii}] and [N~{\sc ii}] centroids presented in the upper panels. 
To establish a  morpho-kinematic parallelism between the centroids obtained and the disk-jet scenario described by this author, the color code was modified. 
Centroids with system velocities from $-100$ to $+100$ km~s$^{-1}$, in the range of the molecular material at the disk, are represented as filled green points, while centroids below $-100$ and above $+100$ km~s$^{-1}$ are shown in blue and red, respectively. 
The color code reveals that those centroids of the [O~{\sc iii}] emission line presumed to belong to the disk are actually aligned along a distinct direction, PA=60$^\circ$, while those at higher velocities deviate from this orientation.

This pattern is more evident in Fig.~\ref{fig:rotated_centroids}, where the position-velocity diagrams of the centroids are displayed along two distinct directions: PA=60$^\circ$ (parallel to the disk) and PA=150$^\circ$ (perpendicular to it). A dashed horizontal line has been added at the position of the CSPN and a vertical line for $V_\mathrm{sys}=+103$ km~s$^{-1}$.
In the direction of PA=60$^\circ$ (upper panel), it can be observed that, within the velocity range of $-100$ to $+140$ km~s$^{-1}$ in the LSR system, the centroids remain relatively flat around zero. 
For velocities below $-140$ and above $+160$ km~s$^{-1}$, there is only a minimal displacement of 0\farcs2 relative to the position of the CSPN.

\begin{figure}
\centering
\includegraphics[width=1\columnwidth]{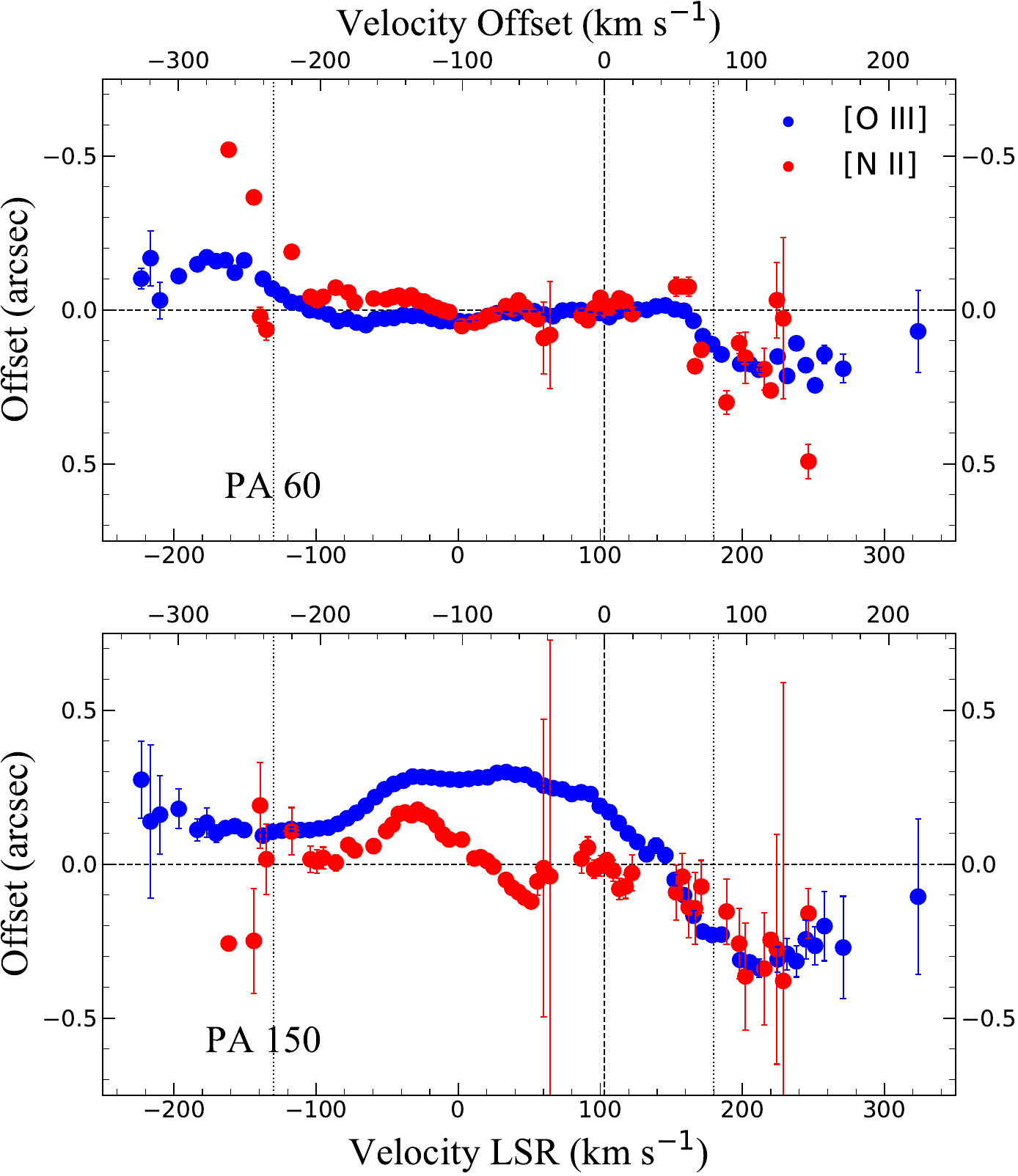}
\caption{
Position-velocity (PV) plots in the [O~{\sc iii}] $\lambda$5007 (red dots) and [N~{\sc ii}] $\lambda$6584 (blue dots) of the CSPN of A\,58 along the PA=60$^\circ$ and PA=150$^\circ$. Positions have been obtained using a 2-dimensional Gaussian fit of the emission at each spectral channel (see text for details). The x-axis are showed in LSR system (bot) and respect to the systemic velocity (top) for both panels.
The systemic velocity of the ejecta and the location of the CSPN are marked by vertical and horizontal dashed lines, respectively. 
}
\label{fig:rotated_centroids}
\end{figure}

\section{Discussion}
\label{sec:discussion}

\subsection{Physical Structure of the H-poor Ejecta of A\,58}
\label{subsec:model}

\begin{figure*}
\begin{center}
\includegraphics[width=1.25\columnwidth]{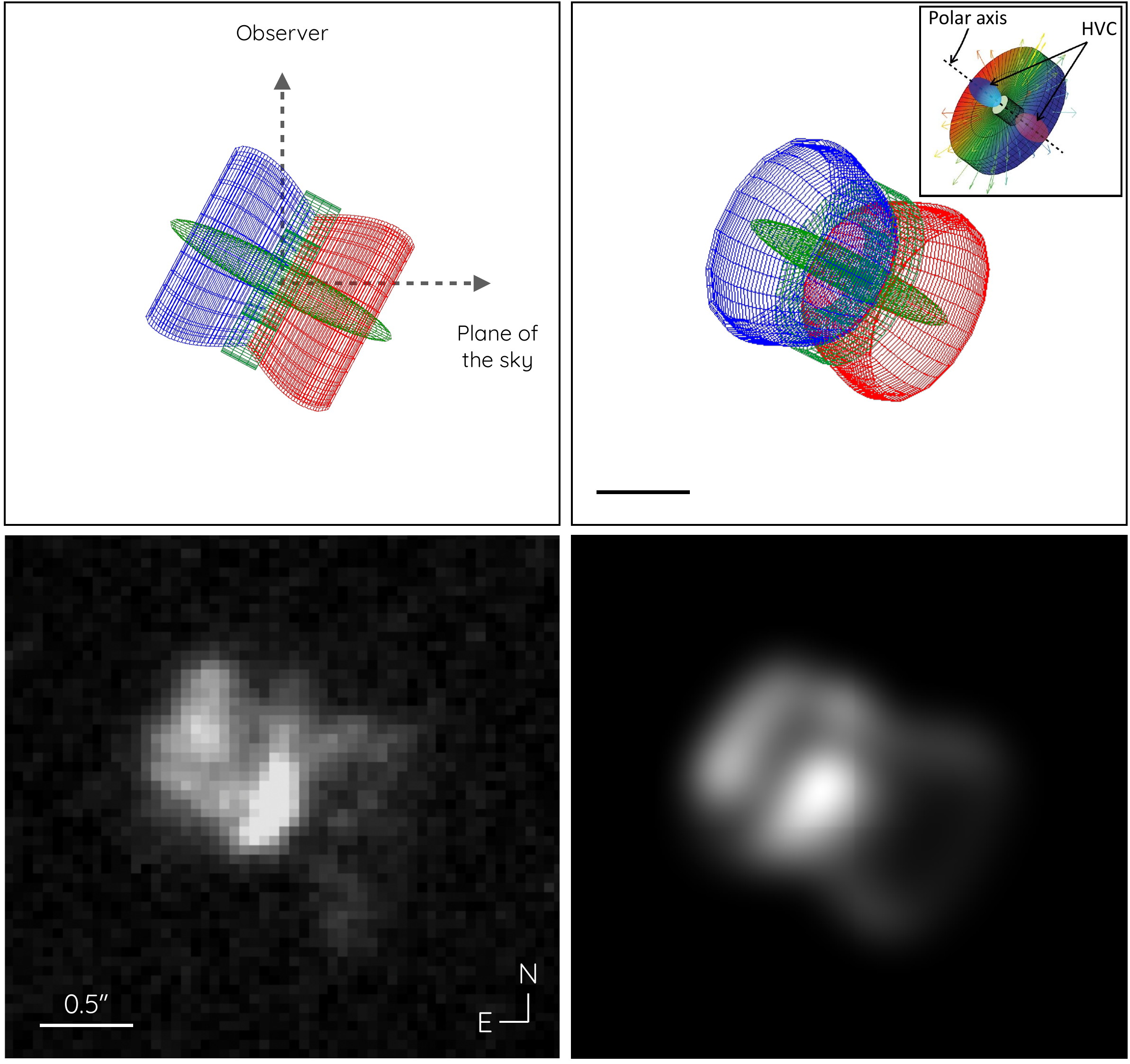}
\caption{
SHAPE mesh model of V605\,Aql and comparison with its \emph{HST} F502N image of the [O~{\sc iii}] $\lambda$5007 \AA\ emission line. 
(top-left) 
Edge-on view of the SHAPE mesh model consisting of a molecular disk (green) and collimated outflows and ionized bipolar outflows (blue and red).  
The inclination angle of the disk with the plane of the sky is 63$^\circ$, as derived from the best-fit model. 
(top-right) 
Projection onto the plane of the sky of the SHAPE mesh model.
For comparison the right panel of Fig.~5 in \cite{Tafoya2022} is shown in the upper right corner of this panel.
(bottom) 
\emph{HST} [O~{\sc iii}] image (left) and synthetic image (right). 
The [O~{\sc iii}] synthetic image is computed from the ionized bipolar outflows.  
The molecular disk does not contribute to the emission, but it rather absorbs the optical emission behind it, which is accounted for adopting an ad hoc emission law for the ionized material 
See Section~\ref{subsec:model} for details.
}
\label{fig:model}
\end{center}
\end{figure*}

The physical structure of the H-poor ejecta of A\,58 has been subject of different studies in the framework of the born-again phenomenon. 
\citet{Pollacco1992}, \citet{Guerrero1996}, and \citet{Clayton2013} investigated it from its optical emission, but these studies are hampered by the high and spatially inhomogeneous extinction.  
Otherwise the radio and sub-mm emission is much less prone to absorption \citep{Tafoya2017,Tafoya2022}.

\citet{Tafoya2022} proposed a model for the physical structure of the molecular gas based on the high-resolution CO~(J=3$\rightarrow$2) emission detected with ALMA. 
The model consisted of a radially expanding disk-like or toroidal structure tilted along a PA of 63$^\circ$ with an inclination angle $i \approx 60^\circ$ with the plane of the sky and an expansion velocity up to 80 km s$^{-1}$, and two high-velocity components (HVCs) orthogonal to the disk with velocities $-140 \lesssim V_{\mathrm{offset}} \lesssim -100$ and $+100 \gtrsim V_{\mathrm{offset}} \gtrsim +140$. 

According to this model based on the CO~(J=3$\rightarrow$2) spatio-kinematic information, the redshifted emission from the disk is found towards the NE of the CSPN, while the blueshifted emission is found towards the SW.  
This is exactly the opposite to the spatial distribution of the [O~{\sc iii}] and [N~{\sc ii}] emission (Fig.~\ref{fig:centroids}), thus indicating that the molecular disk is not emitting in the optical range, or at least, it is not the dominant structure.  
The spatial orientation of the CO HVCs is otherwise consistent with that of the [O~{\sc iii}] and [N~{\sc ii}] emission lines, but the CO HVCs exhibit noticeably lower velocities, in the range from $-140$ to $+140$ km~s$^{-1}$, to those revealed by the optical emission lines (see Fig.~\ref{fig:lines_profiles}), in the velocity range from $-300$ up to $+200$ km~s$^{-1}$. 
The detailed spatial distributions of the CO HVCs and optical emission lines neither agree, with the redshifted optical emission being at the edge of the disk, further away from the CSPN than the corresponding CO HVC, whereas the blueshifted optical emission is generally located closer to the central star than the HVC. 
The velocity and spatial differences between the molecular and ionized components of the ejecta of A\,58 suggest that they do not share the same physical structure.

Elaborating on the high-dispersion spatially-unresolved spectra presented by \citet{Pollacco1992} and using high-resolution multi-epoch \emph{HST} WFPC2 images, \citet{Clayton2013} proposed a sketch for the physical structure of the optical ejecta of A\,58.  
This model consisted of a central disk-like or torus and material moving away from the CSPN perpendicular to the disk, very alike the model proposed for the molecular component by \citet{Tafoya2022}. 
This structure was further geometrically simplified, assuming that the ejecta behaved as an expanding sphere and adopting an extinction that depended on the azimuthal coordinate within it with respect to the line of sight. Although the model reasonably reproduces the emission profile of the spectra, it does not provide information on the actual morphology of the ejecta.
The variations in the MEGARA centroids of [O~{\sc iii}] and [N~{\sc ii}]  (Fig.~\ref{fig:centroids}) do not follow that expected from a simple spherical model, although \citet{Clayton2013} noted variations in the profiles of these emission lines that attributed to small-scale dust inhomogeneities or material clumps.  
These indentations in the emission line profiles are otherwise not present in our higher-quality spectra (Fig.~\ref{fig:lines_profiles}). 
Finally, the symmetric high-velocity ejecta arising from the innermost regions of A\,58 sketched by \citet{Clayton2013} requires the spatial location of blue- and red-shifted velocity channels with similar offset velocity to be distributed symmetrically at both sides of the CSPN, with the systemic velocity at its position, which is not the case (Fig.~\ref{fig:centroids}).

The discrepancies of the predictions of the model presented by \citet{Clayton2013} with the MEGARA observations and the lack of correspondence between the molecular and ionized gas indicate that the physical structure of the latter is rather complex.  
We here propose a model, whose schematic representation is shown in Figure~\ref{fig:model}, that includes the different molecular and ionized gas components of the H-poor ejecta of A\,58.
The upper-left panel of this figure shows a spatio-kinematic model created with the software SHAPE \citep{Steffen2011} consisting of a bipolar structure and a radially expanding disk inclined at an angle of $i=63^{\circ}\pm4^{\circ}$ with respect to the line of sight.  
These structures, oriented along a PA of 63$^{\circ}$ \citep[as proposed by][]{Tafoya2022}, produce the image shown in the upper-right panel of the figure.  
The mesh model shows the combination of the optical (blue and red) and molecular (green) emissions of A\,58, where the geometrical model of the innermost molecular component presented by \citet{Tafoya2022} is shown for comparison.
In this new approach, the optical outflows exhibit an hourglass-like morphology, instead of the high-velocity clumps used for the molecular gas.  
The bipolar lobes have a noticeable thickness, similar to the structure recently proposed for the young molecular ejecta of the Sakurai's object by \citet{Tafoya2023}. 
The high-velocity components observed in the CO~($J=3\rightarrow2$) emission line are located within the cavities formed by the ionized gas, which is escaping around it, most likely transferring momentum to the molecular material and dragging it outwards.

The molecular disk is not expected to emit in the optical emission line, but it rather extincts the emission behind it.  
In this sense it should be noted that SHAPE does not account for radiative transfer physics, thus the effects of the spatially-varying extinction caused by the disk cannot be modeled.  
Instead an arbitrary emission law has been included in the model to simulate both the effects of extinction (emission decreasing with the optical depth along the line of sight and very particularly behind the molecular disk) and distance to the CSPN (emission decreasing with radial distance to the CSPN).  
The bottom panels show the \emph{HST} [O~{\sc iii}] image (left) and a synthetic image rendered from the SHAPE model (right).  
We remark that the bright ``equatorial'' band observed in the \emph{HST} images does not arise from the molecular disk, but it originates from the region of the hourglass structures closest to the disk, where they overlap.

Considering this physical structure for the ionized material and assuming a radial velocity vector with homologous expansion at each point, it is possible to interpret the behavior of the [O~{\sc iii}] emission centroids shown in Fig.~\ref{fig:centroids}.
The emission from the outflow, whose kinematics is best seen in the view from the plane of the sky shown in the left panel of Fig.~\ref{fig:model}, can be split into two components: one expanding close to the plane of the sky (1) and another one expanding mostly perpendicularly to it (2).
The first component would correspond to those centroids of the emission in the low-velocity range from $-100$ to $+100$ km~s$^{-1}$. 
The spatial distribution of the emission at velocities within this range would be reasonably aligned with the direction of the outflow (PA of 63$^\circ$) and would be more intense in the central regions for the blue emission, where material thickness is maximal, and beyond the molecular disk for the red emission, where extinction is minimal. 
These emissions would delineate the maximum spatial extent of the outflow. 
On the other hand, the emission expanding away on near- and far-side within the second, mostly along the line of sight component, will have the highest projected velocities.  
It would be located closer to the CSPN star in the plane of the sky compared to the emission from the first component. 
This behavior is reflected in the centroids of gas with velocities $>+120$ and $<-120$ km~s$^{-1}$. 
We note that the traces of the centroids form a kind of loop, more evident in the less extincted emission heading towards us. 
This may suggest either density inhomogeneities within the structure or emission from the edge of the hourglass structure.

The proposed morphology for the H-deficient ionized and molecular components reveals an evolving physical structure, where the acceleration of material close to the CSPN by its sudden ionization and by the emerging stellar wind generates dynamical effects on the molecular material which has prevailed until now, eroding the molecular disk and accelerating the bipolar molecular outflows.   
In addition to a high adf value in A\,58, in the range $\sim$90 \citep{Wesson2018}\footnote{See \url{https://nebulousresearch.org/adfs/}}, these characteristics support previous suggestions of the presence of a companion in born-again PNe \citep{Soker1997}. 
A companion is definitely required to explain the formation of an equatorial disk and bipolar outflows, but the details of the (stellar or sub-stellar) companion and its orbital parameters are yet unknown. 

It is interesting to note that the abundances of the H-deficient material in A\,58 obtained from optical spectroscopy agree with predictions from nova events and, thus, a nova-like event has been proposed to have taken place here \citep{Lau2011}.
In such models the companion is of stellar origin, but we note that the velocity of the bipolar ejection in A\,58 might actually suggest differently. 
The velocity of the molecular jet must be similar to the escape velocity ($v_\mathrm{esc}$) of the companion and, thus, the $\approx280$~km~s$^{-1}$ velocity of the molecular emission \citep{Tafoya2022} seems to suggest the presence of a substellar object more massive than a Jupiter-like planet ($v_\mathrm{esc}\approx60$~km~s$^{-1}$) or a main-sequence stellar companion with spectral type later than M9 \citep[$v_\mathrm{esc}\approx600$ km~s$^{-1}$;][]{Kaltenegger2009}.

An alternative channel for the formation of disk/jet structures in born-again pNe has recently been proposed by \citet{RodriguezGonzalez2022}.  
It is suggested that a WD in a binary system, after experiencing a VLTP, inflates its outer layers and enters a common envelope with its companion (that, in the case of A\,58, would be a substellar companion). 
Even though the duration of the VLTP is short \citep[$\lesssim$200~yr;][]{Miller2006}, these authors argue that it is sufficient to allow the companion to shape the H-deficient ejecta into a disk plus a bipolar structure.
We note that such claims will have to be put to test with future numerical simulations following the specific evolution of a star experiencing a VLTP in a binary system.

\subsection{Spatial distribution of the HCN and CO high-velocity components}

\begin{figure*}
\centering
\includegraphics[width=1\linewidth]{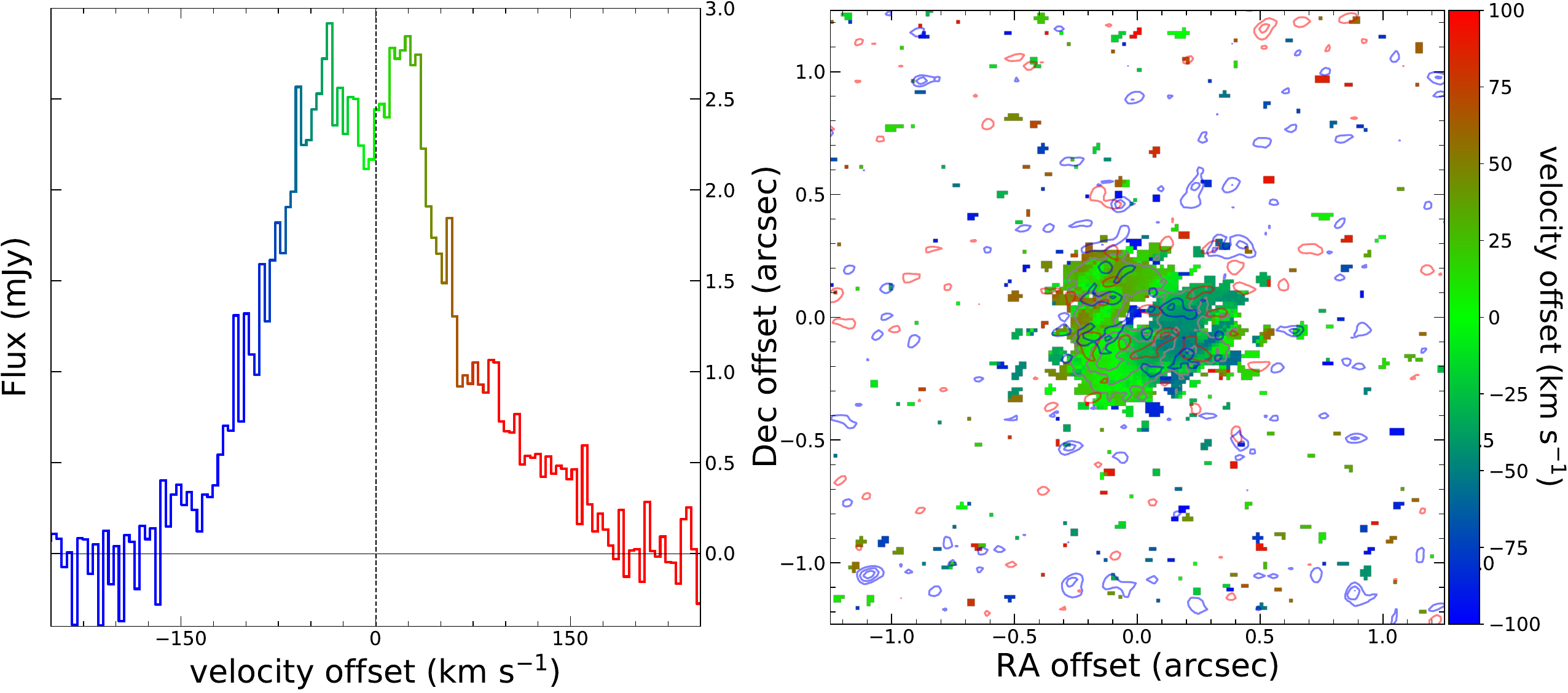}
\caption{Spatio-kinematical distribution of the HCN($J = 4 \rightarrow 3$) emission around V605\,Aql. 
Left: ALMA HCN ($J=4 \rightarrow 3$) emission line profile from V605\,Aql.
Right:
velocity field (first moment) of the HCN ($J=4 \rightarrow 3$) emission around V605\,Aql in the velocity range $-100 < V_\mathrm{offset}$ (km s$^{-1}$) $<$100. is shown as a color map, according to the color code of the emission line profile in the left panel. The pixels with emission below 3 times the rms noise level of 0.2 mJy beam$^{-1}$ were masked. The gray contours show the velocity-integrated emission (zeroth moment) of the HCN ($J = 4 \rightarrow 3$) in the velocity range $-100 < V_\mathrm{offset}$ (km s$^{-1}$) $<$ 100. The gray contours are drawn from $3 \sigma$ on steps of $ 3 \sigma$ (where $\sigma$ = 6.5 mJy beam$^{-1}$~km~s$^{-1}$ is the rms noise level of the zeroth moment image). The blue and red contours show the zeroth moment emission of the HCN ($J = 4 \rightarrow 3$) in the velocity range $-165 < V_\mathrm{offset}$ (km~s$^{-1}$) $< -100$ and $+100 > V_\mathrm{offset}$ (km~s$^{-1}$) $> +$180, respectively. The red contours are drawn for 9, 12 and 15 mJy whereas blue contours are drawn for 7, 9 and 11 mJy.
}
\label{fig:hcn_m1}
\end{figure*}

The ALMA CO molecular emission of A\,58 has been attributed to a radially expanding disk-like structure and orthogonal compact high-velocity components \citep{Tafoya2022}.  
This spatio-kinematical structure differs notably from that derived here from optical emission lines of ionized material.  
Interestingly the spectral profiles of emission lines of other molecules arising from the born-again ejecta, most notably the HCN~(J=$4\rightarrow3$) shown in the left panel of Figure~\ref{fig:hcn_m1}, also present high-velocity components whose spatial distribution can be compared to those of CO~(J=$3\rightarrow2$), [O~{\sc iii}] and [N~{\sc ii}].

The ALMA observations have then been used to obtain the first moment image of the HCN~(J=$4\rightarrow3$) emission line shown in the right panel of Figure~\ref{fig:hcn_m1}.  
Contrary to the CO molecule, the spatial distribution of the high-velocity component of HCN is not compact, but it is rather diffuse and more extended. While CO is well known for being an excellent tracer of molecular outflows, which could include a wide range of physical conditions, HCN and HCO$^{+}$ molecules are typically excited within high-density gas. 
The optical emission lines from ionized material and the HCN molecule would then trace highly excited gas, expelled from regions closer to the central source, whereas the CO molecule would probe denser material from the expanding equatorial disk and molecular outflow.

\section{Conclusions}
\label{sec:conclusions}

We presented the analysis of IFS observations of the born-again PN A\,58 obtained with GTC MEGARA. These observations helped us disclose the true kinematics of the optical emission from the born-again ejecta in this PNe. 
MEGARA's high spectral resolution has allowed us to directly detect the H-alpha emission from the H-poor ejecta without any contamination, which, until now, had only been estimable through various methods, for instance, \citet{Wesson2008} and \citet{Montoro2022}. 
Only the blue component is detected, whereas the red component is deemed to be completely absorbed behind a high-density molecular and dusty disk.

We used observations from CTIO to estimate an average systemic velocity of $+103$ km~s$^{-1}$ consistent with the MEGARA data. 
Previous measurements, both 96 km~s$^{-1}$ obtained by \citet{Clayton2013} and \citet{Tafoya2017} for the inner ejecta, are also consistent within the errors values. 
Nevertheless, the brightening of one side of A\,58 and the blue components of the nebular emission lines being broader than the red ones seem to suggest the interaction of the PN with the ISM, which may result in different radial velocities of the old nebula and recent VLTP ejecta.

We have also performed a channel-by-channel Gaussian adjusting of the ejecta emission, improving the modest spatial resolution of our data. This has allowed us to refine the description of the morphology of the optical emission of the ejecta, which initially was assumed to be practically identical to the molecular emission. The high-velocity components now result in a hourglass structure, within which the high-velocity molecular components are located. The SHAPE model presented reasonably reflects the emission detected in the [O~{\sc iii}] $\lambda$5007 line obtained with the {\it HST}. 
It is worth mentioning that the sketches presented by \citet{Montoro2022} are in good agreement both with the spectral variability observed in the ejecta and with the proposed spatio-kinematic structure.

Prior knowledge of the spatial distribution of the molecular content of the H-poor ejecta of A\,58 has allowed us to interpret the results obtained in the optical range, which would have otherwise been nearly impossible. 
This insight reveals that molecular material may be playing a significant role in the rest of known born-again PNe, namely A\,30, A\,78, HuBi\,1, and the Sakurai's Object.

\begin{acknowledgements}
The authors thanks the referee, Geoffrey Clayton, for his comments and suggestions to improve the presentation of the manuscript. 
B.M.M.\ and M.A.G.\ are funded by grants PGC2018-102184-B-I00 and PID2022-142925NB-I00 of the Ministerio de Educaci\'{o}n, Innovaci\'{o}n y Universidades (MCIU) cofunded with FEDER funds. 
J.A.T. thanks Direcci\'{o}n General de Asuntos del Personal Acad\'{e}mico (DGAPA) of the Universidad Nacional Aut\'{o}noma de M\'{e}xico (UNAM, Mexico) project IA101622, and support from the Marcos Moshinksy Foundation and the Visiting-Incoming programme of the IAA-CSIC through the Centro de Excelencia Severo Ochoa (Spain).
This work as made extensive use of the NASA's Astrophysics Data System (ADS).
\end{acknowledgements}

\bibliographystyle{aa} 
\bibliography{bibliography}

\end{document}